\documentclass[preprint,showpacs,preprintnumbers,amsmath,amssymb]{revtex4}

\usepackage{graphicx}
\usepackage{dcolumn}
\usepackage{bm}

\begin{document}

\title{Temporal Dark Solitons in Nonuniform Bose-Einstein Condensates}
\author{ T.\ Hong, Y.\ Z.\ Wang and Y.\ S.\ Huo}
\address{ Joint Laboratory for Quantum Optics, Shanghai Institute of Optics and Fine Mechanics, Academia Sinica, \\ P.O.Box 800-211,  Shanghai 201800, China }

\begin{abstract}
      We discuss temporal dark solitons in confined 
nonuniform Bose condensates. As a kind of localized high 
excitations, these solitons can be viewed as macroscopic 
quasiparticles, having relative motion to the background 
condensate. We get analytic expression for one dark soliton 
under slowly varying approximation and discuss its special 
propagation properties in nonuniform condensate, then we 
numerically prove that this approximation is reasonable and this 
kind of solitons exhibit their propagation properties in the 
nonuniform condensate. Finally, we simulate the generation of 
dark-soliton-like pulses in the condensate, and indicate that the 
excitation experiment, done by W.~Ketterle's group~\cite{MITSound},
can also be interpreted in terms of temporal dark soliton. 
\end{abstract}
\pacs{ 03.75.Fi, 42.65.Tg, 42.81.Dp }
\date{Received 3 October 1997 by Phys. Rev. A Vol.58, No.4}
\maketitle 

  Bose Einstein condensation(BEC) has been observed in
dilute alkali atomic vapors~\cite{JILABEC,MITBEC,RICEBEC}. 
Recently, further experiments 
have also demonstrated that the condensate can be perfectly 
described by the Gross-Pitaevskii 
equation~\cite{JILAExp,MITExp,Theories,PRuprecht}. For a ground-state
condensate, the kinetic energy in this equation is usually 
negligible as compared with the s-wave scattering interaction 
energy that is the cubic nonlinear term. This indicates that the 
scattering term dominates main properties of the ground-state 
condensate, although it is a weakly interacting system. But for 
an excited condensate in which macroscopic number of atoms 
are collectively excited from ground state into high modes, the 
kinetic energy may become comparable to the scattering 
interaction energy. It could be noticed that the Gross-Pitaevskii 
equation, except for the confining potential, takes the form of  
the cubic nonlinear Schr\"{o}dinger equation (CNLSE) which is 
well-known for the existence of soliton solutions~\cite{ZakDSoliton}. 
Naturally, one will ask whether solitons can exist in a condensate
or not. In the previous works, A.~Mysyrowicz et al. have already found the 
soliton-like propagation of the condensed excitons in $Cu_{2}O$ 
crystal~\cite{AMysyrowicz}, which obeys a similar Gross-Pitaevskii equation 
predicted by E.~Hanamura~\cite{EHanamura}. Weiping Zhang et al., G.~Lenz et 
al., and P.~A.~Ruprecht et al. have proved the existence of bright 
soliton in coherent atomic waves~\cite{WZhang,GLenz} 
and condensate~\cite{PRuprecht} respectively. S.~A.~Morgan et al. 
have analyzed the solitary wave, 
synchronously moving with the background condensate~\cite{SMorgan}.  
The purpose of this paper is to discuss temporal dark solitons, as 
a kind of macroscopic quasiparticles, which have relative 
motion to the background condensate, and it is different from 
the work in reference~\cite{SMorgan}. This kind of solitons is actually 
localized high-mode excitations in the condensate consisting of 
atoms with positive scattering length, such as $^{87}Rb$ or $^{23}Na$. 
First, we will consider a single temporal dark soliton in an idealized 
uniform condensate. Second, we will consider it in a nonuniform 
condensate and discuss its specific propagation features in this 
case. Third, we will numerically simulate the propagation and 
collision of this kind of solitons in nonuniform condensate in 
order to prove that the dark solitons can exist in the nonuniform 
condensate steadily, exhibiting those propagation features, and the approximation adopted in the 
analytical solution is reasonable. Finally, we will also simulate 
the generation of dark-solition-like pulses by arbitrary 
perturbations, using the dark solitons' nonlinear superposition 
properties, and indicate that the excitation experiment, done by 
W.~Ketterle's group~\cite{MITSound}, can also be interpreted in terms of 
temporal dark soliton.

   For a highly excited condensate in which macroscopic 
number of atoms are excited into high excited modes, the 
creation and destruction operators of these modes can be 
considered to be able to commute with each other~\cite{GDMahan}. 
Therefore, in an idealized uniform condensate, its atomic field 
operator can be treated as scalar, called macroscopic wave 
function. The macroscopic wave function can be described by 
the following Gross-Pitaevskii equation~\cite{NoziPine},
\begin{equation}
i\hbar\frac{\partial\Psi\left({\bf r},t\right)}{\partial t}
=-\frac{\hbar^2}{2m}\nabla^2
\Psi\left({\bf r},t\right)
+V_{\rm exc}
\Psi\left({\bf r},t\right)
+U_{0}
\left|
\Psi\left( {\bf r},t\right)
\right|^{2}
\Psi\left({\bf r},t\right)
\label{Gpequationc}
\end{equation}
where $\Psi \left({\bf r},t\right)$  is the macroscopic 
wave function of the 
condensate, including not only the ground sate but also the 
macroscopically populated high excited modes. $V_{exc}$  is an 
external flat potential. $U_{0}=4\pi\hbar^{2}a_{sc}/m$  
is a scattering constant, where 
$a_{sc}$  is the s-wave scattering length and $m$ is the atomic mass. 
Eq.\ (\ref{Gpequationc}) is a CNLSE that possesses soliton solutions.  
When $U_{0}>0$  for positive scattering length, the solutions should 
be nonlinear superposition of dark solitons, according to the 
soliton theorems~\cite{ZakDSoliton}. With the inverse scattering transform 
method~\cite{ZakDSoliton}, we can solve this equation and get one-soliton 
solution,
\begin{equation}
\Psi\left({\bf r}, t\right)
=\Psi_{0}\frac{1+\left(\mu_{s}-i\nu_{s}\right)^{2}e^{-2\Gamma}}
{1+e^{-2\Gamma}}
e^{-i\left(V_{exc}+U_{0}\left|\Psi_{0}
\right|^{2}\right)t/\hbar}
\label{Solitonc}
\end{equation}
where $\Psi_{0}$  is the background amplitude 
of $\Psi\left({\bf r}, t\right)$,  $\mu_{s}$  and $\nu_{s}$ 
are real constants, satisfying $\mu_{s}^{2}+\nu_{s}^{2}=1$,
and $\mu_{s}$  is called the 
eigenvalue of this dark soliton. In this equation, 
\begin{equation}
\Gamma=\nu_{s}\left[\sqrt{U_{0}\left|\Psi_{0}\right|^{2}
m/\hbar^{2}}{\bf k}_{s}\cdot
\left({\bf r}-{\bf r}_{0}\right)+\mu_{s}U_{0}
\left|\Psi_{0}\right|^{2}t/\hbar\right]
\label{Gammac}
\end{equation}
where ${\bf k_{s}}$  is a unit vector in the propagation direction of the 
soliton, ${\bf r_{0}}$  is the center coordinate of 
the soliton. Therefore, the 
atomic number density distribution is
\begin{equation}
\left|\Psi\left({\bf r}, t\right)\right|^{2}
=\left|\Psi_{0}\right|^{2}\left(
1-\nu_{s}^{2}sech^{2}\Gamma\right)
\label{Density}
\end{equation}
where $\nu_{s}^{2}$ is called the darkness of the dark soliton. 
Additionally, it is worthy to be emphasized that dark soliton 
possesses no threshold, so it can be stimulated easily. According 
to the theorems of dark soliton~\cite{SAGredeskul}, 
an arbitrary perturbation of 
the wave function can be described as a nonlinear superposition 
of solitons, which indicates that soliton has its universality in the 
space described by the CNLSE (\ref{Gpequationc}). 
From expression (\ref{Density}), we can 
find that the soliton is a localized function, because when 
$z\rightarrow \pm \infty$ , its dark density decreases 
to zero quickly. Thus we can 
derive its full width at half peak darkness, 
\begin{equation}
\triangle_{\rm s\left(FWHM\right)}
=\frac{arccosh\sqrt{2}}{\nu_{s}\sqrt{\pi a_{sc}
\left|\Psi_{0}\right|^{2}}}
\label{Widthc}
\end{equation}
This expression has a similar form with the healing length of the 
Bose condensate~\cite{EPGross}, except for the characteristic constant 
of the soliton. It is known that a two-dimensional vortex core in 
the Bose condensate has a similar size to the healing 
length. Consequently, their 
common features tell us that the dark soliton could be viewed as 
a kind of one-dimensional vortex core. Although there is no 
circular current around this kind of core due to its noncircularly-connected 
feature, the both sides of the soliton are connected by 
its bottom current density, which we will give more discussion 
later. Similar to a two-dimensional vortex, this one-dimensional 
vortex has a `` vortex line'', or more exactly, it may be called 
vortex plane, which is the center plane of the soliton 
perpendicular to the vector ${\bf k_{s}}$. Corresponding to the 
circulation around a vortex line in two-dimensional case, there is 
also a similar quantity for the one-dimensional vortex core, 
which is just the multiply of the width and the velocity of the 
soliton and is directly determined by its characteristic constant. 
However, for a dark soliton, its characteristic constant is usually 
not a quantized, but a continuously valued number, this feature 
indicates that the one-dimensional vortex is not quantized for an 
uniform Bose condensate, and therefore the corresponding one-dimensional
 flow should exhibit classical fluidity instead of 
superfluidity. But, this does not mean that it won't become 
quantized for the externally confined nonuniform Bose 
condensate, because in considering the solitonic nonlinear 
dynamics at the vicinity of the confined boundaries, this 
problem always becomes very complicated and couldn't be 
solved analytically at present. With Eq.\ (\ref{Widthc}), we can 
calculate that when the atomic number density of the condensate 
$\left|\Psi_{0}\right|^{2}$  is $10^{14}cm^{-3}$ , 
the width of a black ($\nu_{s}=1$)  soliton is $0.95\mu m$ 
for condensate of $^{23}Na$. Obviously, it can be much smaller than 
the real size of a nonuniform condensate, for example, $17\mu m$  in 
radial direction and $300\mu m$  in axial direction, realized by 
W.~Ketterle's group~\cite{MITBEC}. So the previous idealized uniform 
condensate is reasonable for very dark solitons.

  Subsequently, we consider a more real and special case in 
which the condensate, confined in external potential, is 
nonuniform, and it has a shape of cigar. Similar to the above 
assumption, a macroscopic number of atoms are still assumed to 
be excited, and the commutation relations of the atomic field 
operators are still valid. Thus, the Gross-Pitaevskii equation can 
be expressed as
\begin{equation}
i\hbar\frac{\partial\Psi\left({\bf r},t\right)}{\partial t}
=-\frac{\hbar^2}{2m}\nabla^2
\Psi\left({\bf r},t\right)
+V_{\rm ext}\left({\bf r}\right)
\Psi\left({\bf r},t\right)
+U_{0}
\left|
\Psi\left( {\bf r},t\right)
\right|^{2}
\Psi\left({\bf r},t\right)
\label{GPequation}
\end{equation}
where $V_{\rm ext}\left({\bf r}\right)
=V_{r}\left({\bf R}\right)+V_{a}\left(z\right)$ 
 is the external confining potential, 
which can be divided into two parts, the radial
 $V_{r}\left({\bf R}\right)$  and the 
axial $V_{a}\left( z \right)$  in cylindrical coordinates. 
Additionally, We assume 
the wave function $\Psi\left({\bf r}, t\right)$ 
 can be divided into the background 
ground-state condensate $\Phi \left( {\bf r}\right)$
 and a soliton function $\beta\left(z, t\right)$  
propagating along $z$ axis,
\begin{equation}
\Psi\left({\bf r},t\right)
=\beta\left( z,t\right)
\Phi\left({\bf r}\right)
e^{-i\mu_{c}t/\hbar}
\label{Seperate}
\end{equation}
where $\mu_{c}$  is the chemical potential
of the stationary ground-state 
of the condensate. Then Eq.\ (\ref{GPequation}) 
is transformed into

\begin{eqnarray}
\left[
i\hbar\frac{\partial \beta\left( z,t\right)}{\partial t}
+ \frac{\hbar^{2}}{2m}
\frac{\partial^{2}\beta\left(z,t\right)}{\partial z^{2}}
+\frac{\hbar^{2}}{m}
\frac{\partial\beta\left(z,t\right)}{\partial z}
\frac{\partial ln\Phi\left({\bf r}\right)}{\partial z}
\right]
\Phi\left({\bf r}\right)
\nonumber \\
+\left[
-  U_{0}\left|\beta\left( z,t\right)\Phi\left({\bf r}\right)
\right|^{2}\beta\left(z,t\right)
+U_{0}\left|\Phi\left({\bf r}\right)\right|^{2}
\beta\left(z,t\right)
\right]
\Phi\left({\bf r}\right) 
\nonumber \\
=  \beta\left(z, t\right)\left[
-\mu_{c}
\Phi\left({\bf r}\right)
-\frac{\hbar^{2}}{2m}
\nabla^{2}\Phi\left({\bf r}
\right)
+V_{ext}\left({\bf r}\right)
\Phi\left({\bf r}\right)
+U_{0}\left|
\Phi\left({\bf r}\right)\right|^{2}
\Phi\left({\bf r}\right)\right]
\label{ExGr}
\end{eqnarray}
Because $\Phi\left({\bf r}\right)$  is the 
stationary ground-state wave function of 
the condensate, it must satisfy
\begin{equation}
\mu_{c}
\Phi\left({\bf r}\right)
=-\frac{\hbar^{2}}{2m}
\nabla^{2}\Phi\left({\bf r}
\right)+V_{ext}\left({\bf r}\right)
\Phi\left({\bf r}\right)
+U_{0}\left|
\Phi\left({\bf r}\right)\right|^{2}
\Phi\left({\bf r}\right)
\label{groundstate}
\end{equation}
Consequently, Eq.\ (\ref{ExGr}) is reduced to
\begin{eqnarray}
i\hbar\frac{\partial \beta\left( z,t\right)}{\partial t}
+\frac{\hbar^{2}}{2m}
\frac{\partial^{2}\beta\left(z,t\right)}{\partial z^{2}}
+\frac{\hbar^{2}}{m}
\frac{\partial\beta\left(z,t\right)}{\partial z}
\frac{\partial ln\Phi\left({\bf r}\right)}{\partial z}
\nonumber \\
-U_{0}\left|\beta\left( z,t\right)\Phi\left({\bf r}\right)
\right|^{2}\beta\left(z,t\right)
+U_{0}\left|\Phi\left({\bf r}\right)\right|^{2}
\beta\left(z,t\right)
=0
\label{Excitation}
\end{eqnarray}
Furthermore, we assume 
$\left|\Phi\left({\bf r}\right)\right|^{2}$
  is a slowly varying function of 
coordinate $z$, as compared with $\beta\left(z, t\right)$ ,
 satisfying
\begin{equation}
\frac{1}{2}\left|
\frac{\partial^{2}\beta\left(z,t\right)}
{\partial z ^{2}}\right|
\gg
\left|\frac{\partial\beta\left( z, t\right)}{\partial z}
\frac{\partial ln\Phi \left({\bf r}\right)}{\partial z}
\right|
\label{Slowvary}
\end{equation}
Therefore, we can ignore the third term in
 Eq.\ (\ref{Excitation}) , and get
\begin{equation}
i\hbar\frac{\partial\beta\left( z,t\right)}{\partial t}
+\frac{\hbar^{2}}{2m}
\frac{\partial^{2}\beta\left( z,t\right)}{\partial z^{2}}
-U_{0}
\left|
\beta\left(z,t\right)
\Phi\left( {\bf r}\right)
\right|^{2}
\beta\left(z,t\right)
+U_{0}
\left|
\Phi\left( {\bf r}\right)
\right|^{2}
\beta\left(z,t\right)
=0
\label{Excitation2}
\end{equation}
Because $\left|\Phi\left({\bf r}\right)\right|^{2}$  
is a slowly varying function in contrast to a 
soliton, when we solve this equation with inverse scattering 
transform method for a single soliton solution, we can take 
$\left|\Phi\left({\bf r}\right)\right|^{2}$  as a constant. 
Subsequently, we can routinely get one 
soliton expression,
\begin{equation}
\beta\left(z, t\right)
=\frac{1+\left(\mu_{s}-i\nu_{s}\right)^{2}e^{-2\Gamma}}
{1+e^{-2\Gamma}}
\label{Soliton}
\end{equation}
where
\begin{equation}
\Gamma=\nu_{s}\left[
\sqrt{U_{0}\left|\Phi\left({\bf r}\right)
\right|^{2}m/\hbar^{2}}{\bf k}_{s}\cdot
\left({\bf r}-{\bf r}_{0}\right)+\mu_{s}U_{0}
\left|\Phi\left({\bf r}\right)\right|^{2}t/\hbar\right]
\label{Gamma}
\end{equation}
 $\mu_{s}$ and $\nu_{s}$  still follow 
the previous definition. $z_{0}$  is the center 
coordinate of this soliton. 
The full width at half peak darkness of 
the soliton is
\begin{equation}
\triangle_{\rm s\left(FWHM\right)}
=\frac{arccosh\sqrt{2}}{\nu_{s}\sqrt{\pi a_{sc}
\left|\Phi\left({\bf r}\right)\right|^{2}}}
\label{Width}
\end{equation}
which is different from Eq.\ (\ref{Widthc}) 
for $\left|\Phi\left({\bf r}\right)\right|^{2}$
being a slowly varying 
function of  $z$. This means 
$\triangle_{\rm s\left(FWHM\right)}$   
is to be increased when the 
soliton moves from the ``top'' to the ``downhill'' of the ground 
condensate wave function.  From Eq.\ (\ref{Gamma}) we can derive the 
velocity of the soliton,
\begin{equation}
{\bf v}\left({\bf r}\right)
=-\mu_{s}\sqrt{\frac{U_{0}\left|
\Phi\left({\bf r}\right)\right|^{2}}{m}}{\bf k}
\label{Velocity}
\end{equation}
where ${\bf k}$  is the unit vector in 
the direction of $z$. ${\bf v}\left({\bf r}\right)$ varies 
with $\Phi\left({\bf r}\right)$, 
which is consistent with the local speed of sound 
given by Bogoliubov~\cite{NBogoliubov} 
and Lee, Huang, and Yang~\cite{LeeHuangYang}. 
Additionally, Eq.\ (\ref{Velocity})  
tells us that the velocity of this 
temporal dark soliton is also relevant to the eigenvalue, and it 
is, more explicitly, usually less than the absolute value of the 
corresponding sound speed in the nonuniform Bose 
condensate. However, this property can not be gotten in the 
density perturbation theory
 of hydrodynamics, such as ref.\ \cite{EZaremba}. 
In considering the local density dependent property of the 
velocity, we can find that there is a slow feedback process 
between the velocity and the displacement of the soliton, and 
the temporal dark soliton behaves as an oscillator in the 
nonuniform Bose condensate~\cite{THong}.
 Additionally, for those radial tightly confined 
Bose condensates, the velocity of the soliton varies rapidly 
with the density in the radial direction. This leads to very 
serious radial dispersion of the temporal dark soliton, to 
which a similar phenomenon has been observed in the 
experiments done by W.~Ketterle's group~\cite{MITSound}. As it is 
known in nonlinear optics, the $1+1$ dimensional dark solitons 
are usually instable under their transverse perturbations, and 
always evolve into dark vortices~\cite{BLDavies}. The 
analogy between this solitonic excitation in Bose condensate 
and the optical dark solitons enlighten us that the temporal 
dark soliton must be instable under the stretch of transverse 
dispersion, and it may also evolve into vortices. Due to the 
cylindrical symmetry of the ground state condensate, it is 
more likely to evolve into vortex rings~\cite{NoziPine}, because 
they can still preserve the cylindrical symmetry. The 
circulation of those stable vortex rings must be quantized, 
which is a direct consequence of the coherent property of the 
Bose condensate, relevant to the nature of the superfluidity of 
the Bose condensate. Our further work on this topic is still in 
progress. Assuming that the radial variation could be 
neglected for some cases, we will find the product 
\begin{equation}
\triangle_{\rm s\left(FWHM\right)}\cdot v\left({\bf r}\right)
=-2 arccosh\sqrt{2}\frac{\hbar\mu_{s}}{m\nu_{s}}
\label{Constant}
\end{equation}
should be a constant for a darkness-fixed soliton. As we have 
described above, analogous to the circulation around the vortex 
line of a two-dimensional vortex in the fluid, this quantity may 
also be called a one-dimensional circulation, which 
characterizes the properties of the one-dimensional dark soliton 
vortex core directly. But the more important thing is that it is 
very realistic for an experimental measurement of this quantity. 
Therefore, Eq.\ (\ref{Constant}) may offer us a useful method to judge 
whether a dark pulse in the condensate can be very precisely 
interpreted as a temporal dark soliton.
From Eq.\ (\ref{Seperate}) we can also 
get the atomic current density carried 
by the soliton
\begin{equation}
{\bf j}
=-\frac{i\hbar}{2m}\left(
\Psi^{\ast}\left({\bf r},t\right)
\nabla\Psi\left({\bf r},t\right)
-\Psi\left({\bf r},t\right)
\nabla\Psi^{\ast}\left({\bf r},t\right)\right)
={\bf k}\mu_{s}\nu_{s}^{2}U_{0}^{1/2}m^{-1/2}
\Phi^{3}\left({\bf r}\right)sech^{2}\Gamma
\label{Current}
\end{equation}
which is pulsed. The width of this pulse is proportional to that of 
the soliton. It is worthy to be noticed that the direction of the 
current density pulse is opposite to the propagation direction of 
the dark soliton, however, the propagation direction and the 
velocity of the current density pulse are both same to those of 
the soliton. When $\nu_{s}=0$  or $\nu_{s}=1$,
 the amplitude of this current 
density pulse reaches its minimum $j_{\rm min}=0$. 
When  $\nu_{s}=2\sqrt{3}/9$, it 
gets to its maximum
\begin{equation}
j_{\rm max}=\frac{4\sqrt{69}}{243}
U_{0}^{1/2}m^{-1/2}\Phi^{3}\left({\bf r}\right)
\label{MCurrent}
\end{equation}
The existence of this maximum shows that each distribution of 
the current density usually corresponds to two possible soliton 
eigenvalues and therefore two one-dimensional dark soliton 
vortices. This property is different from the usual two-
dimensional vortex case. Additionally, from Eq.\ (\ref{Current}), we 
can find that the relationship of the current density and the 
soliton eigenvalue $\mu_{s}$ is in the shape of S. In 
considering some nonlinear dynamical process, such as the 
solitonic excitation or the soliton collision with a boundary. This 
relationship may provide us an opportunity of finding bistable 
phenomena of the temporal dark soliton. 

In the above derivation, we have adopted slowly varying 
approximation (\ref{Slowvary}). To make sure this approximation is 
reasonable and temporal dark solitons can really exhibit 
their propagation properties in a confined nonuniform 
condensate, we numerically simulate the propagation and 
collision of this kind of solitons. We assume the 
ground-state condensate is composed of $N=5\times 10^{6}$
 of  $^{23}Na$  atoms 
with $a_{sc}=27.5\AA$~\cite{ETiesinga}, 
and the radial trapping frequency is 
$\omega_{r}=2\pi\times 1800 Hz$, the axial trapping frequency is 
$\omega_{a}=2\pi\times 18Hz$. During the simulation, we adopt 
one-dimensional Thomas-Fermi approximation in the radial 
direction of the cylindrically symmetric condensate. 
Consequently, we can approximate Eq.\ (\ref{GPequation}) as 
one-dimensional Gross-Pitaevskii equation by integration in the 
radial cross section of the condensate,
\begin{equation}
i\hbar\frac{\partial\psi_{a}\left( z,t\right)}{\partial t}
+\frac{2}{3}\mu_{c}\psi_{a}\left( z,t\right)
=-\frac{\hbar^{2}}{2m}\frac{\partial^{2}\psi_{a}\left(z,t\right)}
{\partial z^{2}}+V_{a}\left(z\right)\psi_{a}\left(z,t\right)
+\frac{2}{3}U_{0}\left|\psi_{a}\left(z,t\right)\right|^{2}
\psi_{a}\left( z, t\right)
\label{Numerical}
\end{equation}
where $\psi_{a}\left(z,t\right)$ is the axial 
wave function of the condensate, 
distributed along the axial direction $z$, 
and $V_{a}\left( z\right)=m\omega_{a}^{2}z^{2}/2$. 
Although the one-dimensional Thomas-Fermi approximation is 
still not strict, it can still keep 
the axial wave function $\psi_{a}\left(z,t\right)$ 
without slowly varying approximation (\ref{Slowvary}) 
in the direction of $z$, 
and therefore Eq.\ (\ref{Numerical}) is sufficient to 
give the approximation a 
test. We numerically solve Eq.\ (\ref{Numerical}) 
by split-Fourier transform 
method~\cite{MEguchi}. As an example, two solitons' 
collision process is 
illustrated in Fig.1. The two soliton, 
$\nu_{s1}=0.5$, $\nu_{s2}=0.5$,  set 
initially on the ``downhill'' of the 
condensate wave function with 
distance $6.2\mu m$, shown in Fig.1.(a). 
It should be noted that both 
the wave function and the 
atomic number density, shown in all 
figures in this paper, have already been transformed into 
dimensionless forms with respect to their ground-state center 
values at $z=0.0\mu m$. These two solitons propagate and collide 
with each other, and then they separate and propagate 
independently, as shown in Fig.1.(b). Before and after the 
collision, they can conserve their shape steadily. From the 
contour graph of this collision, as shown in Fig.1.(c), the widths 
of the solitons vary with their position, which indicates the 
locality of their widths. The velocities of the two solitons are all 
about $v=15\mu m/ms$. There is no damage occurred to both of 
these solitons, so the collision must be elastic. More numerical 
experiments show that so long as the numerical precision is 
enough, the solitons can propagate steadily as far as one can 
compute. Therefore, we can draw a conclusion that the slowly 
varying approximation~(\ref{Slowvary}) is reasonable, and the solitons 
indeed have the propagation properties, indicated above in 
nonuniform condensate confined by external potential. 
Additionally, as we have seen, the properties of these temporal 
dark solitons in the condensate are very similar to those of 
conventional particles, so we can consider them as a kind of 
quasiparticles. But they are different from phonons, because 
every one of them is macroscopic whereas a phonon is usually 
microscopic, and furthermore every soliton must be a coherent 
structure as described in Eq.\ (\ref{Soliton}).

    According to the dark soliton theorems~\cite{ZakDSoliton}, 
dark solitons have nonlinear superposition properties. We may use these 
properties to generate temporal dark solitons in 
nonumiform condensate. To illustrate this possibility, we 
numerically simulate the generation of soliton-like pulses 
by arbitrary perturbations in the nonuniform condensate. 
As an example, here we give two reversed Gaussian 
function perturbations in the condensate, shown in Fig.2.(a). 
The darkness of these two perturbations are $0.44$  and $0.64$ 
respectively, and the full widths of them are $2.47\mu m$ 
and $0.99\mu m$ respectively, and their distance is  $10\mu m$. As 
shown in Fig.2.(b), each of these two dark perturbations 
splits quickly into more pairs of dark pulses, propagating 
with different velocity depending on its own darkness. The 
collision of the two out-splitted dark pulses must be elastic 
because the two pulses can still keep even symmetry with 
their own twins. Obviously, these properties are very 
similar to those of solitons. From Fig.2.(c), which is the 
contour graph of Fig.2.(a), we can find that all the 
velocities of these out-splitted soliton-like pulses are 
around  $V=20\mu m/ms$. Although we can't make sure 
that these out-splitted pulses won't split anymore ( in fact, 
the pair, splited by the left one, have already begun to split 
again into more pairs of pulses), with the above elastic 
collision evidence, we can at least draw a conclusion that 
the initial dark pulses can be approximated as multi-solitons, 
each of which is composed of a number of solitons. 
The recent experiment, done by W.~Ketterle's group~\cite{MITSound}, 
has shown a similar phenomenon as we have simulated 
here. In their experiment, one dark perturbation splits into 
two, and then these two out-splitted pulses continue to 
propagate with spreading widths. Although there is no 
description about observation of further splitting of the out-split 
dark pulse, the out-split dark pulse must be composed 
of more than one solitons. The spreading of the pulses can 
be interpreted as the radial dispersion as we have described 
above, but one can't exclude some other possible reasons, 
such as further splitting of the multi-solitons and the 
widening with the decrement of the background condensate 
as described by Eq.~(\ref{Width}). And it can't be usually simply 
explained only as the last one, except that one can, by 
chance, get twin strict dark solitons, every one of which 
must have the form as we have described analytically in 
Eq.\ (\ref{Soliton}). However, further splitting may not be observable, 
because the size of the condensate is limited, the velocities 
differences are all to decrease to zero due to the Eq.\ (\ref{Velocity}), 
and they may therefore have no enough distance to 
propagate for further splitting. To test the soliton-like 
properties of the dark pulses in the present experiment, 
according to our numerical simulation, we suggest that one 
can generate two dark pulses propagating in opposite 
direction, and observe their collision. One can also test 
these properties by measuring the relationship between the 
velocity of every steadily propagating out-split dark pulse 
and its width, and Eq.\ (\ref{Constant}) can be used as a 
judgment. Certainly, more precise interpretation of the 
experiment would need more study in details and more 
realistic simulation which can't be included in this short 
paper. But, as we have briefly discussed above, the 
temporal dark soliton theorems are promising tools to 
interpret the experiment done by W.~Ketterle's group~\cite{MITSound}.

   In conclusion, we have derived the temporal dark soliton in 
nonuniform condensate under slowly varying approximation. 
Viewed as the one-dimensional vortex cores, these temporal 
solitons may provide deeper insight into the superfluidity in the 
cigar-shaped Bose-Einstein condensate. Then we have 
numerically simulated the propagation and collision of this kind 
of solitons, and proved that the slowly varying approximation is 
reasonable and the solitons can exhibit the special propagation 
properties in the nonuniform condensate. We have finally 
simulated the generation of soliton-like pulses by two arbitrary 
dark pulses in the nonuniform condensate, and with this 
simulation, we have given the experimental results of 
W.~Ketterle's group a qualitative interpretation in terms of  
temporal dark soliton.

We would like to thank Dr. Jaren Liu, Dr. Xueru Zhang, Dr. 
Xunming Liu, Dr. Xinqi Wang, Dr. Fusheng Li, and Mr. 
Wenbao Wang for their fruitful discussions and helps. 
This work is supported by National Science Foundation of 
China, under Grant No.19392503.

\newpage

\begin{figure}[h]
\includegraphics[width=10cm, height=5cm, angle=0]{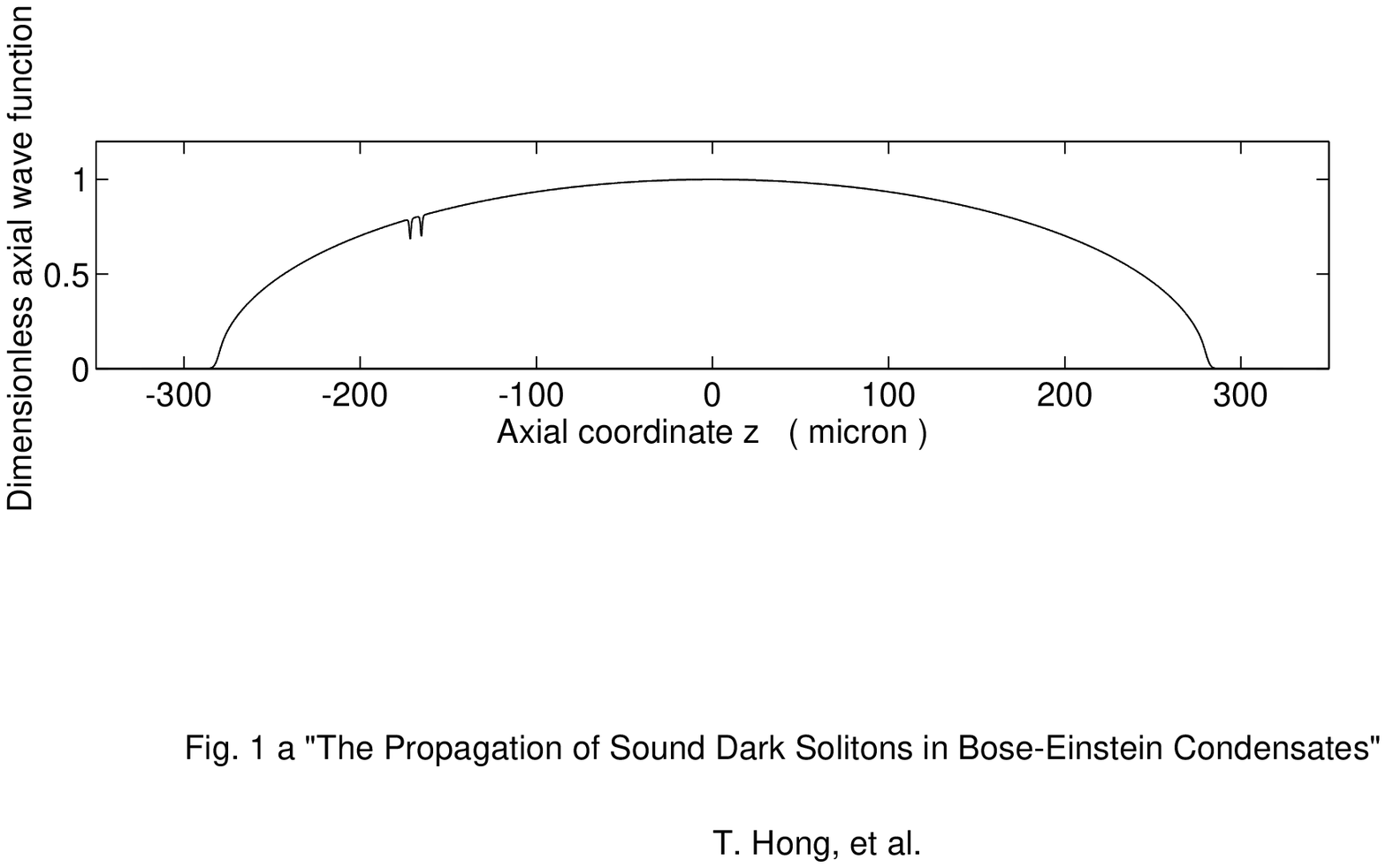}
\includegraphics[width=10cm, height=5cm, angle=0]{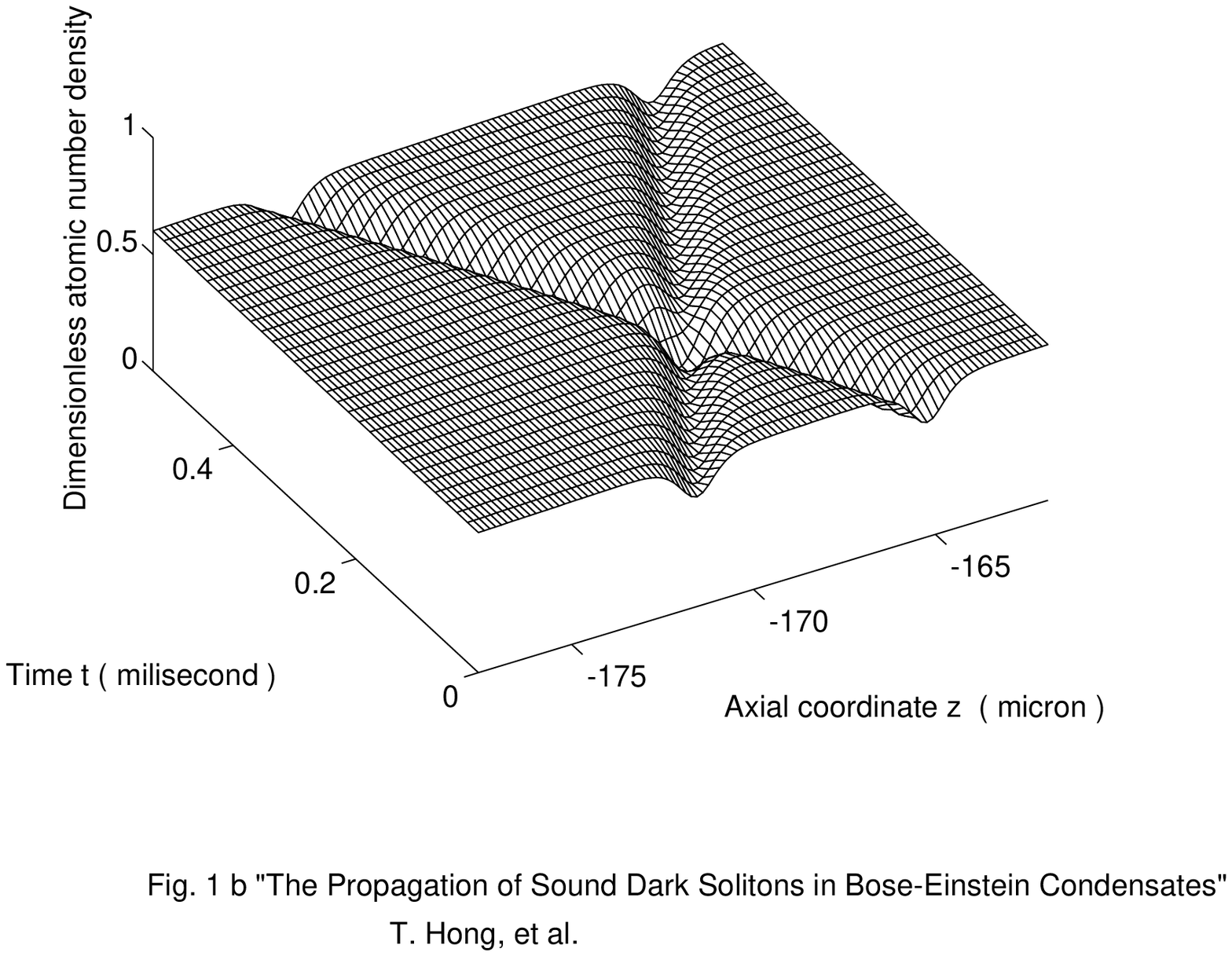}
\includegraphics[width=10cm, height=5cm, angle=0]{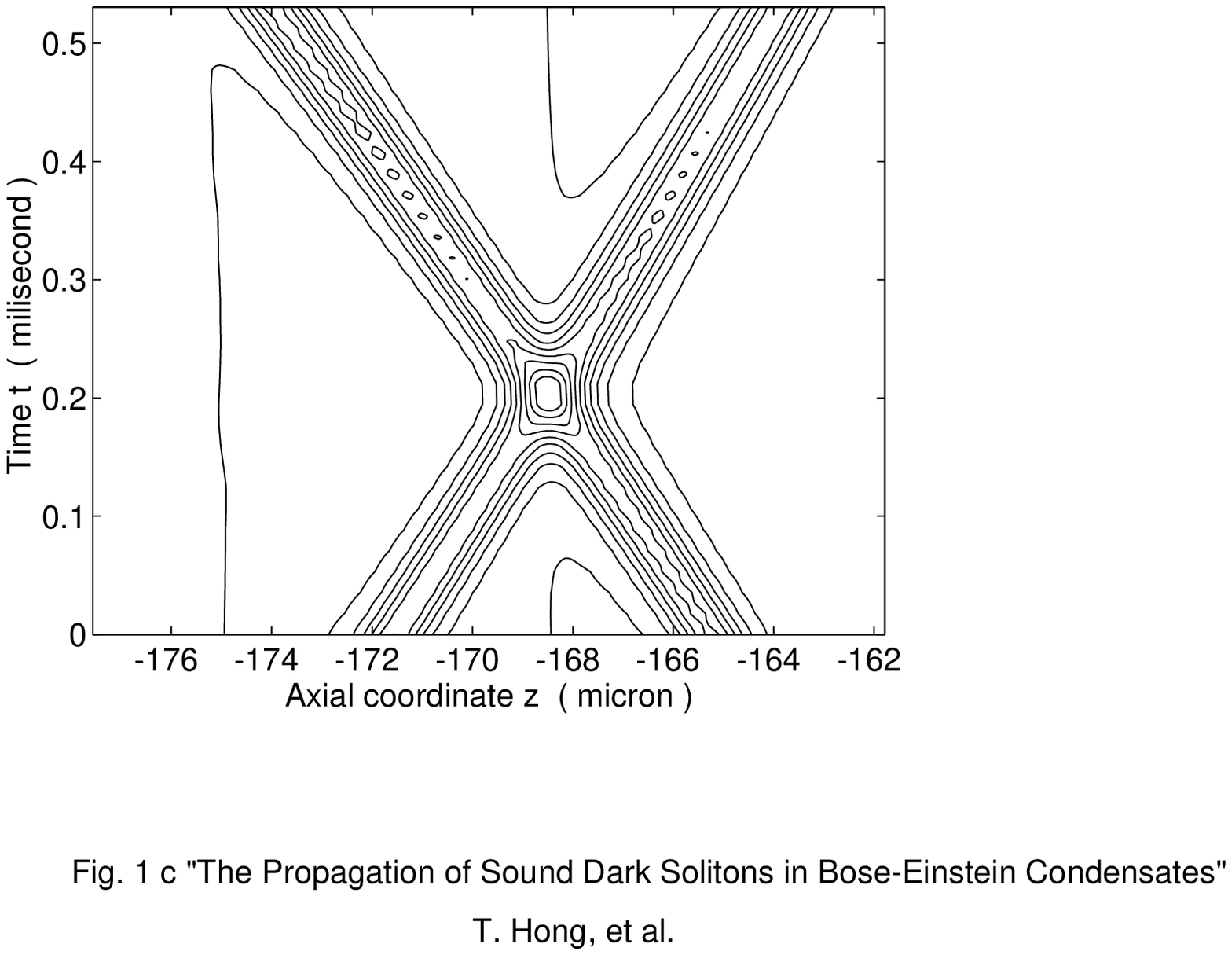}
\caption{ The collision of two temporal dark solitons in the 
condensate. (a) The dimensionless axial wave function of the ground-
state condensate is initially excited with two dark solitons. (b) The two temporal dark solitons propagate steadily and collide with each other elastically. (c) The contour graph of the two colliding solitons. Both 
the widths of them vary with their position on the condensate.}
\label{fig1}
\end{figure}

\begin{figure}[h]
\includegraphics[width=10cm, height=5cm, angle=0]{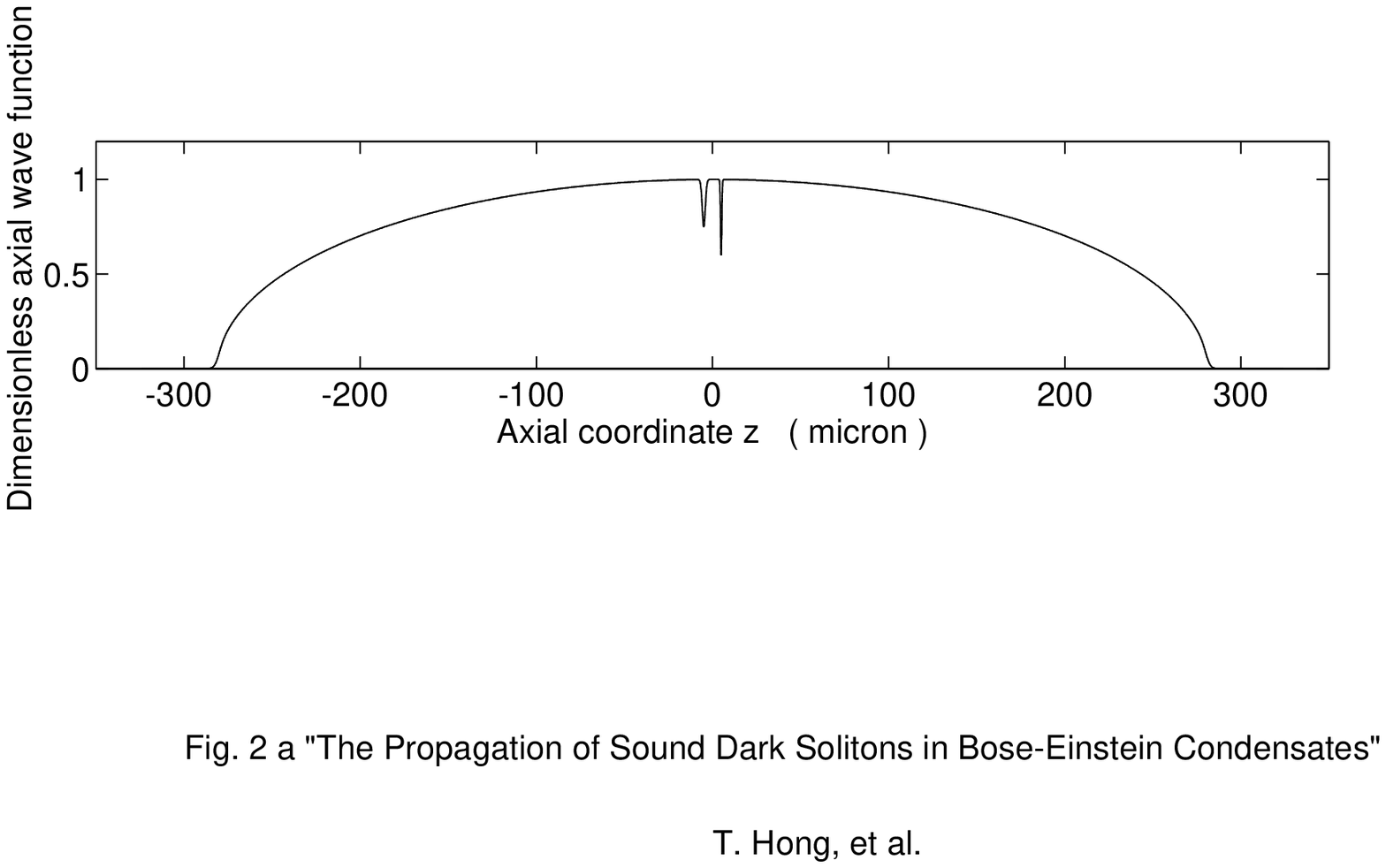}
\includegraphics[width=10cm, height=5cm, angle=0]{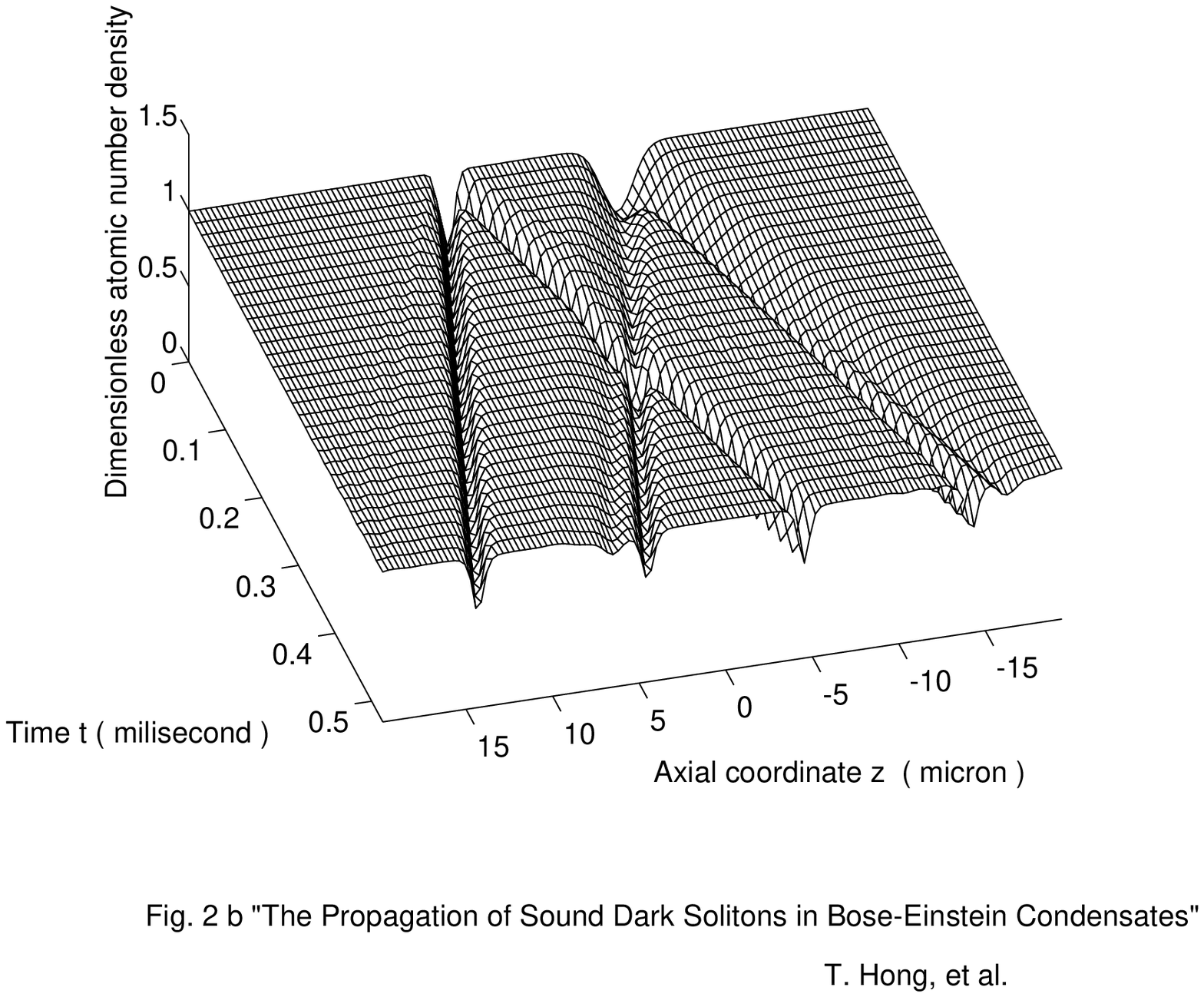}
\includegraphics[width=10cm, height=5cm, angle=0]{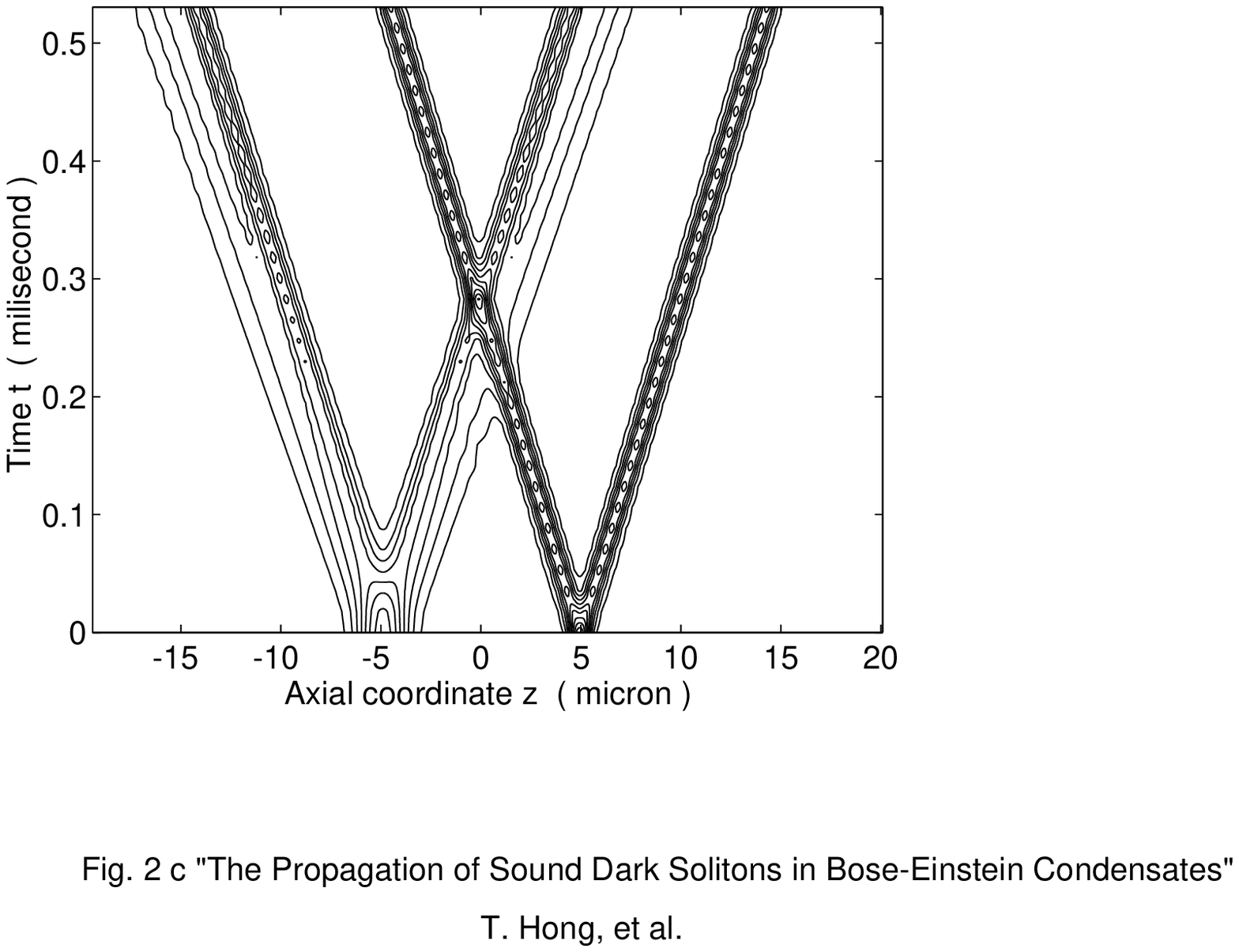}
\caption{ The excitation of dark temporal dark soliton in a ground-
state condensate.\\
      (a) The dimensionless axial wave function of the ground-
state condensate is initially excited with two reversed Gaussian 
shape dark pulses.\\
      (b) Each of two dark pulses splits into pairs of dark 
pulses, and the split pulses collide with each other elastically. 
This indicates that the initial dark pulses can be approximated as 
nonlinear superposition of temporal dark solitons. \\ 
      (c) The contour graph of the two dark pulses. Some of 
the split dark pulses propagate steadily and some of them are splitting into more dark pulses.}
\label{fig1}
\end{figure}


\newpage
\begin{references}
\bibitem{JILABEC} M.\ H.\ Anderson, J.\ R.\ Ensher, M.\ R.\ Matthews, 
C.\ E.\ Wieman, and E.\ A.\ Cornell, Science {\bf 269}, 198(1995).
\bibitem{MITBEC} K.\ B.\ Davis, M.-O.\ Mewes, M.\ R.\ Andrews, N.\ J.\ van 
Druten, D.\ S.\ Durfee, D.\ M.\ Kurn, and W.\ Ketterle, Phys.\ Rev.\ Lett. 
{\bf 75}, 3969(1995); M.-O.\ Mewes, M.\ R.\ Andrews, N.\ J.\ van Druten, 
D.\ M.\ Kurn, D.\ S.\ Durfee, and W.\ Ketterle, Phys.\ Rev.\ Lett. {\bf 77}, 
416(1996).
\bibitem{RICEBEC} C.\ C.\ Bradley, C.\ A.\ Sackett, J.\ J.\ Tollett, and 
R.\ G.\ Hulet, Phys.\ Rev.\ Lett. {\bf 75}, 1687(1995);  C.\ C.\ Bradley, 
C.\ A.\ Sackett, and R.\ G.\ Hulet, Phys.\ Rev.\ Lett. {\bf 78}, 985(1997).
\bibitem{JILAExp} D.\ S.\ Jin, J.\ R.\ Ensher, M.\ R.\ Matthews, 
C.\ E.\ Wieman, and E.\ A.\ Cornell, Phys.\ Rev.\ Lett. {\bf 77}, 420(1996); 
D.\ S.\ Jin, M.\ R.\ Matthews, J.\ R.\ Ensher, C.\ E.\ Wieman, and 
E.\ A.\ Cornell, Phys.\ Rev.\ Lett. {\bf 78}, 764(1997).
\bibitem{MITExp} M.-O.\ Mewes, M.\ R.\ Andrews, N.\ J.\ van Druten, 
D.\ M.\ Kurn, D.\ S.\ Durfee, C.\ G.\ Townsend, and W.\ Ketterle, 
Phys.\ Rev.\ Lett. {\bf 77}, 988(1996).
\bibitem{Theories} M.\ Edwards, P.\ A.\ Ruprecht and K.\ Burnett, 
R.\ J.\ Dodd and C.\ W.\ Clark, Phys.\ Rev.\ Lett. {\bf 77}, 1671(1996); 
A.\ L.\ Fetter, Phys.\ Rev.\ A, {\bf 53}, 4245(1996);
V.\ M.\ P\'{e}rez-Carc\'{i}a, H.\ Michinel, J.\ I.\ Cirac,
M.\ Lewenstein, and P.\ Zoller, Phys.\ Rev.\ Lett. {\bf 77}, 5320(1996);
K.\ G.\ Singh and D.\ S.\ Rokshar,
Phys.\ Rev.\ Lett. {\bf 77}, 1667(1996); S.\ Stringari, Phys.\ Rev.\ Lett. 
{\bf 77}, 2360(1996).
\bibitem{PRuprecht} P.\ A.\ Ruprecht, M.\ J.\ Holland, and K.\ Burnett, 
Phys.\ Rev.\ A, {\bf 51}, 4704(1995).
\bibitem{ZakDSoliton} V.\ E.\ Zakharov and A.\ B.\ Shabat, 
Zh.\ Eksp.\ Teor.\ Fiz. {\bf 64}, 1627(1973)[Sov.\ Phys.\ -JETP {\bf 37}, 
823(1973)].
\bibitem{AMysyrowicz} A.\ Mysyrowicz, {\it Bose-Einstein 
Condensation}, edited by A.\ Griffin, D.\ W.\ Snoke, and
S.\ Stringari (Cambridge University Press, Cambridge, UK, 1995).
\bibitem{EHanamura} E.\ Hanamura, Solid State Comm. {\bf 91}, 
889(1994).
\bibitem{WZhang} Weiping Zhang, D.\ F.\ Walls, and B.\ C.\ Sanders, 
Phys.\ Rev.\ Lett. {\bf 72}, 60(1994).
\bibitem{GLenz} G.\ Lenz, P.\ Meystre, and E.\ M.\ Wright, Phys.\ Rev. 
\ Lett. {\bf 71}, 3271(1993).
\bibitem{SMorgan} S.\ A.\ Morgan, R.\ J.\ Ballagh, and K.\ Burnett, 
Phys.\ Rev.\ A, {\bf 55}, 4338(1997).
\bibitem{MITSound} M.\ R.\ Andrews, D.\ M.\ Kurn, H.-J.\ Miesner, 
D.\ S.\ Durfee, C.\ G.\ Townsend, S.\ Inouye, and W.\ Ketterle,  
Phys.\ Rev.\ Lett. {\bf 79} 553(1997).
\bibitem{GDMahan} G.\ D.\ Mahan, {\it Many-Particle Physics} (Plenum Press, New York, 1990).
\bibitem{NoziPine}  P.\ Nozi\`{e}res and D.\ Pines, {\it Theory of 
Quantum Liquids} ( Addison-Wesley, Redwood City, California, 1990 ), {\bf  2.}
\bibitem{SAGredeskul} S.\ A.\ Gredeskul, Yu.\ S.\ Kivshar, and 
M.\ V.\ Yanovskaya, Phys.\ Rev.\ A, {\bf 41}, 3994(1990).
\bibitem{EPGross} E.\ P.\ Gross, J.\ Math.\ Phys.\ {\bf 4}, 195(1963).
\bibitem{NBogoliubov} N.\ Bogoliubov, J.\ Phys. {\bf 11}, 23(1947).
\bibitem{LeeHuangYang} T.\ D.\ Lee, K.\ Huang, and C.\ N.\ Yang, 
Phys.\ Rev. {\bf 106}, 1135(1957).
\bibitem{EZaremba} E.\ Zaremba, Phys.\ Rev.\ A\ {\bf 57}, 518(1998).
\bibitem{THong} T.\ Hong, Y.\ Z.\ Wang, Y.\ S.\ Huo, accepted by Chinese Phys.\ Lett..
\bibitem{BLDavies} B.\ L.-Davies, J.\ Christou, V.\ Tikhonenko, 
Y.\ S.\ Kivshar, J.\ Opt.\ Soc.\ Am.\ B\ {\bf 14}, 3045(1997).
\bibitem{ETiesinga} E.\ Tiesinga et al., 
J.\ Res.\ Natl.\ Inst.\ Stand.\ Technol. {\bf 101}, 505(1996).
\bibitem{MEguchi} Masashi Eguchi, Kazuya Hayata and 
Masanori Koshiba, J.\ Appl.\ Phys. {\bf 72}, 3255(1992).
\end{references}
\end{document}